\documentclass[aps, prd, amsmath, floats, floatfix, twocolumn, nofootinbib, superscriptaddress, showpacs,fleqn]{revtex4-1}
\usepackage{graphicx}
\usepackage{color}
\usepackage{latexsym}
\usepackage{array}
\usepackage{units}
\usepackage[normalem]{ulem}
\usepackage[breaklinks=true,hyperindex=true,
pdftitle={Gravitational-wave versus X-ray tests of strong-field gravity},
colorlinks=true,pagebackref=false,citecolor=blue,plainpages=false,pdfpagelabels,
linkcolor=blue,urlcolor=blue]{hyperref}

\newcommand{\be}{\begin{eqnarray}}
\newcommand{\ee}{\end{eqnarray}}

\newcommand{\GR}{{\mbox{\tiny GR}}}
\newcommand{\LO}{{\mbox{\tiny 0PN}}}
\newcommand{\GW}{{\mbox{\tiny GW}}}
\newcommand{\RZ}{{\mbox{\tiny RZ}}}
\newcommand{\fut}{{\mbox{\tiny Fut}}}
\newcommand{\cata}{{\mbox{\tiny Obs}}}
\newcommand{\cataT}{{\mbox{\tiny GWTC-1}}}

\newcommand{\odos}{{\mbox{\tiny O2}}}
\newcommand{\otres}{{\mbox{\tiny O3a}}}
\newcommand{\ocinco}{{\mbox{\tiny O5}}}

\begin{document}
\title{Gravitational-wave versus X-ray tests of strong-field gravity}

\author{Alejandro~C\'ardenas-Avenda\~no}
\email[Corresponding author: ]{ac54@illinois.edu}
\affiliation{Programa de Matem\'atica, Fundaci\'on Universitaria Konrad Lorenz, 110231 Bogot\'a, Colombia}
\affiliation{Department of Physics, University of Illinois at Urbana-Champaign, Urbana, IL 61801, USA}
\affiliation{eXtreme Gravity Institute, Department of Physics, Montana State University, 59717 Bozeman MT, USA}

\author{ Sourabh Nampalliwar}
\affiliation{Theoretical Astrophysics, Eberhard-Karls Universit\"{a}t T\"{u}bingen, Auf der Morgenstelle 10, 72076 T\"{u}bingen, Germany}

\author{Nicol\'as Yunes}
\affiliation{Department of Physics, University of Illinois at Urbana-Champaign, Urbana, IL 61801, USA}
\affiliation{eXtreme Gravity Institute, Department of Physics, Montana State University, 59717 Bozeman MT, USA}

\date{\today}

\begin{abstract}

Electromagnetic observations of the radiation emitted by an accretion disk around a black hole, as well as gravitational wave observations of coalescing binaries, can be used to probe strong-field gravity.
We here compare the constraints that these two types of observations can impose on theory-agnostic, parametric deviations from the Schwarzschild metric.
On the gravitational wave side, we begin by computing the leading-order deviation to the Hamiltonian of a binary system in a quasi-circular orbit within the post-Newtonian approximation, given a parametric deformation of the Schwarzschild metrics of each binary component. 
We then compute the leading-order deviation to the gravitational waves emitted by such a binary in the frequency domain, assuming purely Einsteinian radiation-reaction. 
We compare this model to the LIGO-Virgo collaboration gravitational wave detections and place constraints on the metric deformation parameters, concluding with an estimate of the constraining power of aLIGO at design sensitivity. 
On the electromagnetic side, we first simulate observations with current and future X-ray instruments of an X-ray binary with a parametrically-deformed Schwarzschild black hole, and we then estimate constraints on the deformation parameters using these observations.
We find that current gravitational wave observations have already placed constraints on the metric deformation parameters than are slightly more stringent than what can be achieved with current X-ray instruments.
Moreover, future gravitational wave observations with aLIGO at design sensitivity by 2026 will be even more stringent, becoming stronger than constraints achievable with future ATHENA X-ray observations before it flies in 2034.
\end{abstract}


\maketitle


\section{Introduction}
Until recently, nearly all of our knowledge about astronomical objects had been obtained through the electromagnetic radiation they produce, or that is generated around them. However, we have recently entered the gravitational wave detection era, which has provided new and interesting data that is shaping our current understanding of fundamental physics~\cite{yunes2016theoretical}. With this new type of observations and with the improvement of existing techniques, we are already learning about theoretical physics in the \emph{extreme gravity} regime, where the curvature of spacetime is large and  the gravitational field is strong and dynamical. 

Although a plethora of precision tests in the Solar System, with binary pulsars, and with cosmological observations have confirmed the predictions of Einstein's theory of General Relativity (GR), this theory is only now being thoroughly tested in the extreme gravity regime~\cite{Will:2014kxa}. An ideal laboratory for testing strong-field gravity is astrophysical black holes (BHs). Bearing in mind that at the moment there is no evidence that such astronomical bodies carry sufficient net electric charge to affect the metric (in particular because of the extreme weakness of gravity relative to electromagnetism~\cite{young1976capture,luminet1998black}), isolated BHs in GR are described by the Kerr metric, as required by the so-called \emph{no hair theorems}~\cite{carter1971axisymmetric,robinson1975uniqueness}. Any observation suggesting otherwise would be an indication of a violation of the axioms of these theorems, which include the possibility of beyond-Einstein physics~\cite{cardoso2019testing}. This program is commonly referred to as \emph{testing the Kerr hypothesis}, and it has been pursued over the past years using electromagnetic observations~\cite{johannsen2010testing,sadeghian2011testing,broderick2014testing,johannsen2016testing,bambi2016testing} and gravitational waves~\cite{gossan2012bayesian,rodriguez2012verifying,meidam2014testing,thrane2017challenges,krishnendu2017testing,isi2019testing}.  

\begin{figure}[]
\includegraphics[width=\linewidth{}]{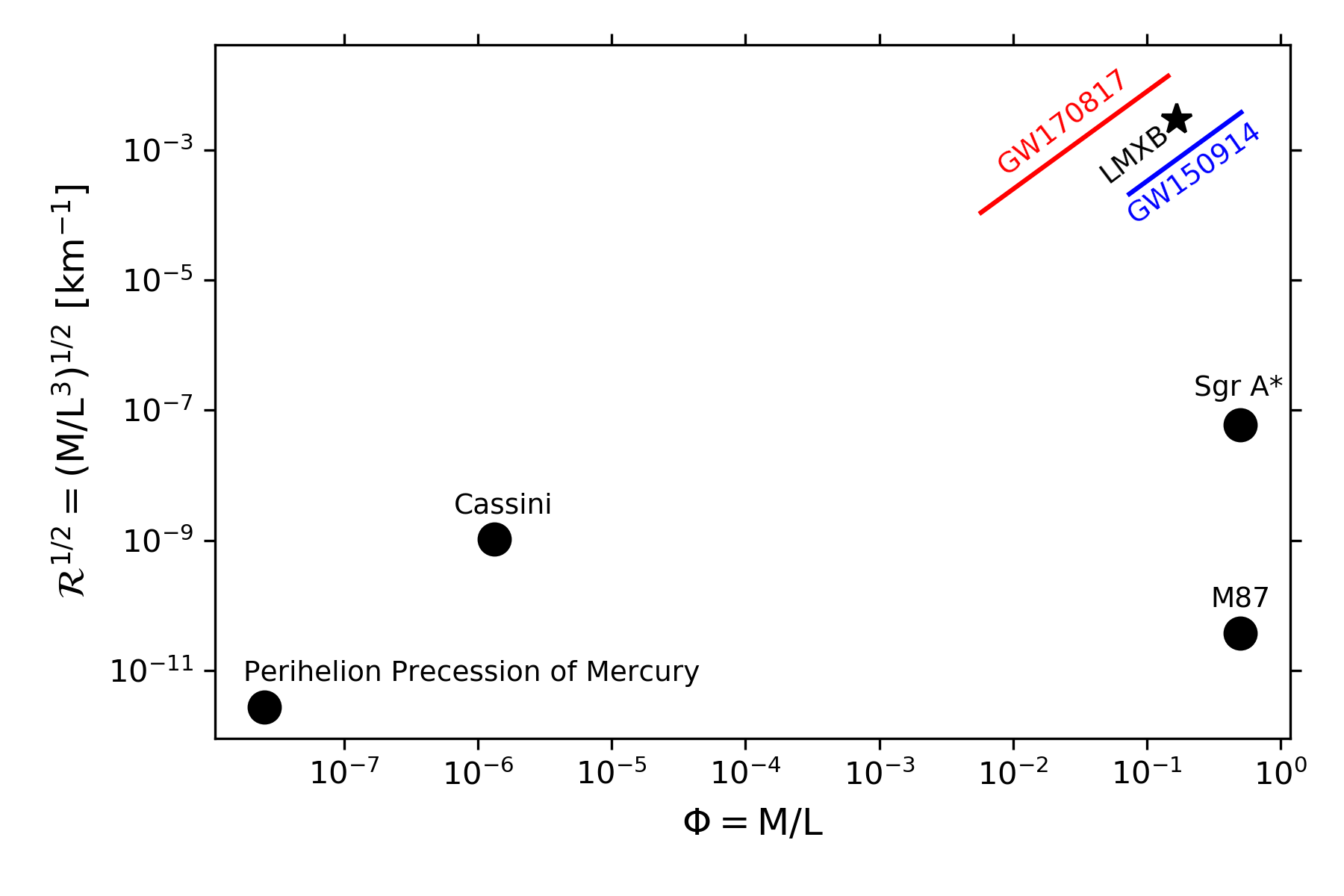}
\caption{Illustrative diagram of the curvature-potential phase space sampled by some experiments that test GR. The vertical axis denotes the square root of the characteristic curvature length scale $ \mathcal{R} = M / L^{3}$, while the horizontal axis the characteristic gravitational potential $\Phi = M/L$. For the GW150914 and GW170817 events, we evaluated $\cal{R}$ and $\Phi$ from $20$ Hz to merger, taking $L$ to be the orbital separation and $M$ the total mass. For the low-mass X-ray binary (LMXB), we used $M\sim10\,M_{\odot}$ and $L\sim6M$ for the location of the innermost-stable circular orbit of a Schwarzschild BH. Observe that gravitational-wave and LMXB observations both have access to the strong-field regime, in the upper-right corner of phase space, while only electromagnetic observations currently have access to large potentials but lower curvatures through the observation of supermassive BHs.}
\label{fig:extreme}
\end{figure}

Placing constraints on (or finding) modifications from GR with data is not an easy task. This is both because deviations may be intrinsically small, and because all modified gravity theories to date lead to the same spacetime behavior far away from the BH, making weak-field tests ineffective. However, modifications to GR may not be so small in the strong-field regime, for instance near the event horizon, where distinctive features may arise. One way to classify and understand different systems in terms of their gravitational strength consists of computing the characteristic curvature $ \mathcal{R} = M / L^{3}$ and the characteristic gravitational potential $\Phi=M/L$, where $M$ and $L$ are the characteristic mass and size of the system, respectively. Following Refs.~\cite{psaltis2008probes,baker2015linking,yunes2016theoretical}, Fig.~\ref{fig:extreme} compares the regions in curvature-potential phase space that are probed by gravitational-wave and X-ray observations, including also for reference the regions probed by the Mercury-Sun system through perihelion precession observations~\cite{Will:2014kxa}, the Cassini satellites~\cite{bertotti2003test}, and the targets of the Event Horizon Telescope, i.e., Sgr A* and M87~\cite{Akiyama:2019cqa}. The strong-field gravity regime is located in the right corner of the phase space of Fig.~\ref{fig:extreme}, where we have current data from different systems and precision tests can be carried out. Observe that current gravitational wave observations with ground-based detectors are limited to this upper-right corner of phase space, while electromagnetic observations have access to the entire right side because the latter can include supermassive BHs.   

Electromagnetic observations of the radiation emitted by X-ray binaries~\cite{remillard2006x}, as well as gravitational wave observations of coalescing binaries~\cite{Yunes:2013dva} are complementary in the following sense~\cite{psaltis2008probes,Will:2014kxa,cardenas2016testing}. Tests using the electromagnetic spectrum probe the non-dynamical configuration of vacuum gravitational fields. These tests rely on observables that use photons and plasma as test particle tracers of the spacetime geometry. On the other hand, gravitational waves tests probe both the conservative and the dissipative sector of a theory, because they rely both on the time-symmetric part of the gravitational fields through the Hamiltonian of the system, and on the dissipative part of the radiation-reaction force built from the field perturbations. It is therefore theoretically possible for electromagnetic observations to be more sensitive to certain aspects of modifications to the conservative sector of a theory than gravitational wave observations by avoiding confusion and correlations from the dissipative sector. Electromagnetic observations, however, are affected by other astrophysical modeling uncertainties~\cite{cardenasexp2019}, which deteriorate its constraints relative to the gravitational wave ones.      

In this work we compare tests of GR in terms of constraints on parametric deformations of the spacetime away from the Schwarzschild metric. 
We model spacetime deformations through the parametrization introduced by Rezzolla \& Zhidenko~\cite{Rezzolla:2014mua}, which includes the Schwarzschild metric when the deformation parameters vanish, while also allowing for a wide range of BH solutions in specific modified gravity theories. The generation of gravitational waves is modified if the spacetime in the neighborhood of the compact objects in a binary system is described by such a Schwarzschild deformation. This is because the post-Newtonian Hamiltonian (or Lagrangian) is constructed from the metric itself, so if the spacetime is modified near either of the compact objects, then the Hamiltonian is also modified. The latter implies the equations of motion are modified, which then affect the evolution of the orbital phase and ultimately of the gravitational waves emitted. 

\begin{figure*}[htb]
	\centering
\includegraphics[width=\columnwidth,clip=true]{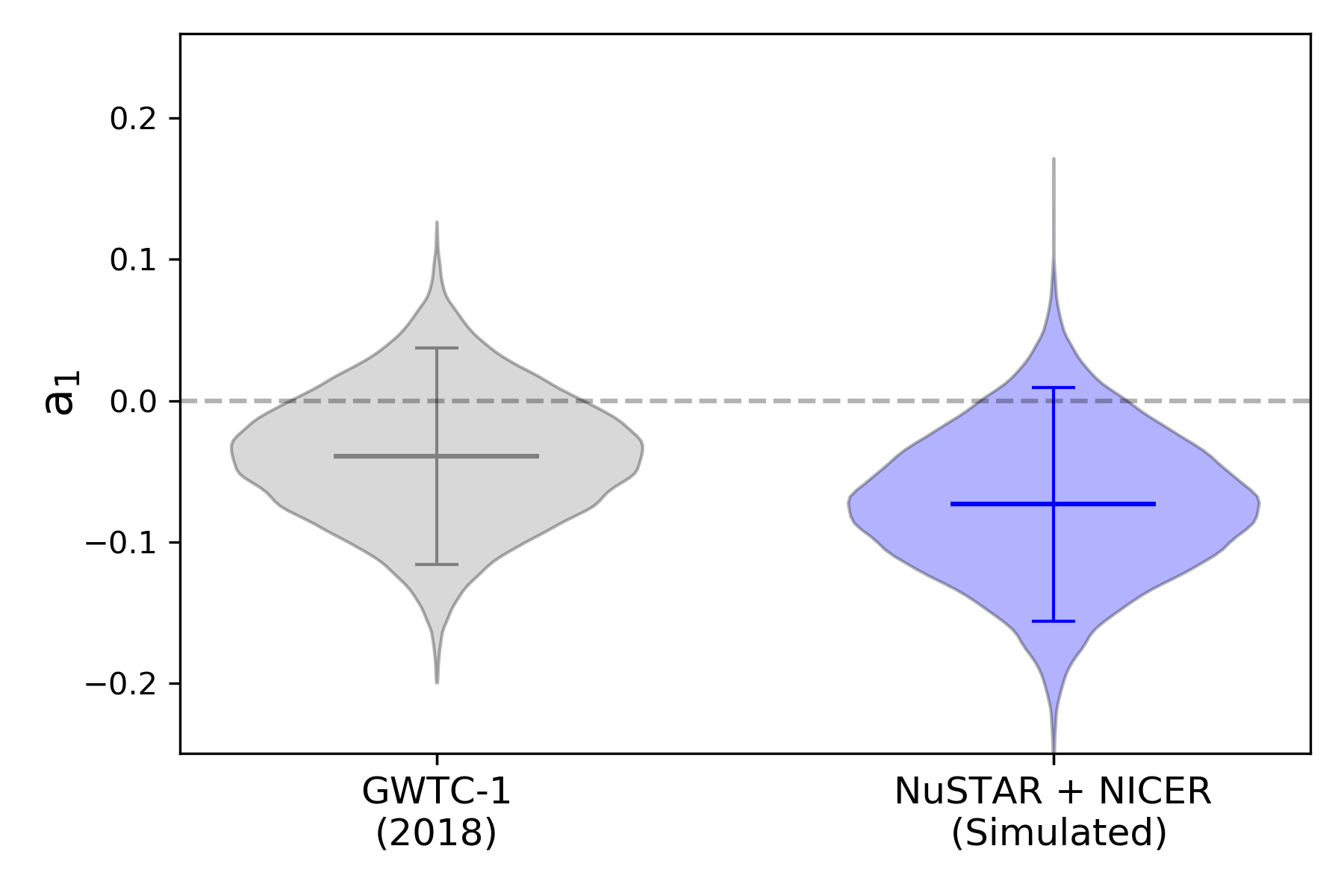}
\includegraphics[width=\columnwidth,clip=true]{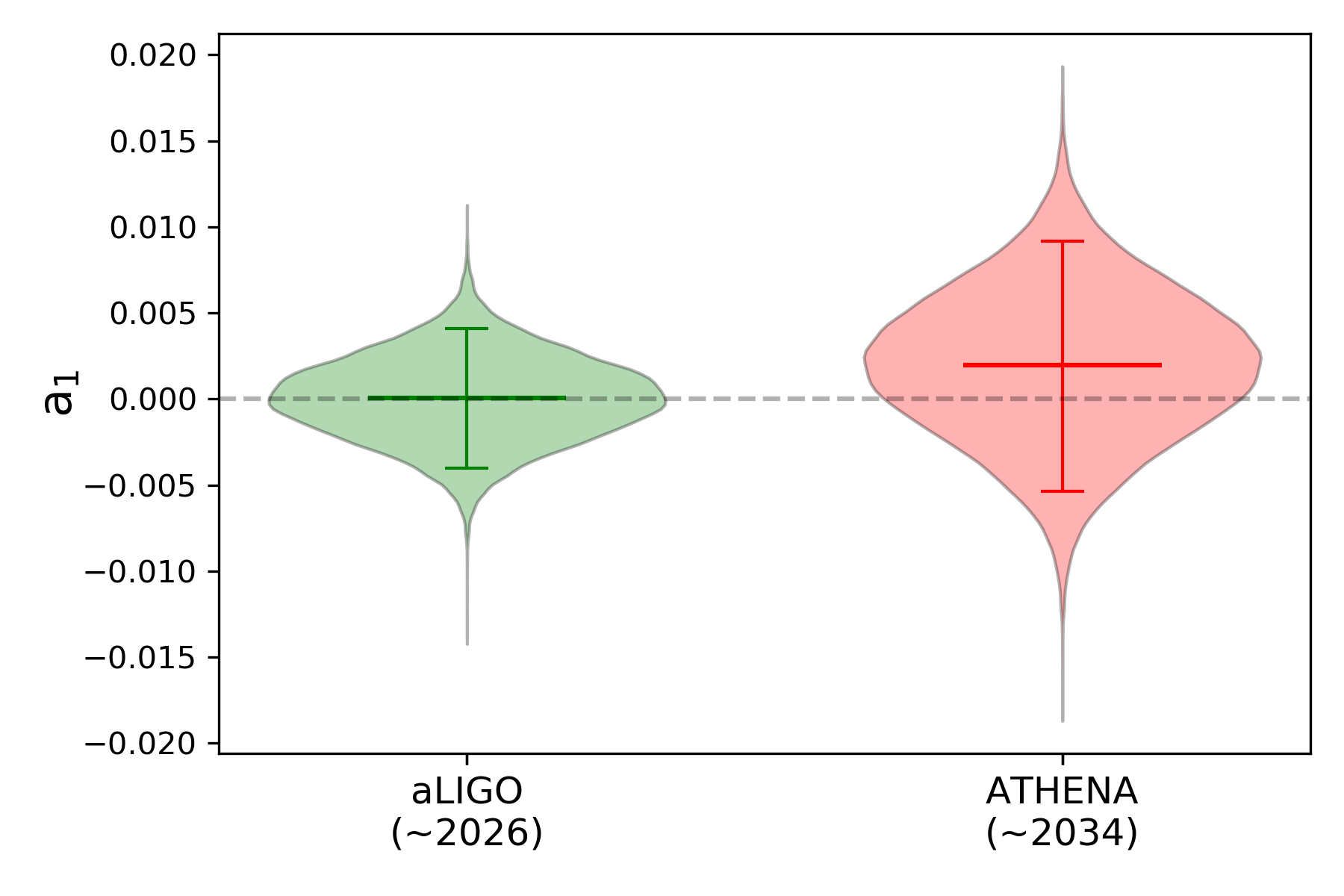}
\caption{Marginalized posterior of the bumpy parameter $a_{1}$ using current (left panel) and future (right panel) gravitational-wave and X-ray observations. Current observations mean here the strongest events reported in the LIGO-Virgo Catalog GWTC-1~\cite{ligo2019tests}, and simulated X-ray data with current instruments, NuSTAR and NICER, while future observations mean those achievable with aLIGO at design sensitivity ($\sim2026$) and simulated observations with ATHENA ($\sim2034$). The short horizontal lines indicate the 90\% credible interval around the indicated mean, while the long dashed horizontal line at zero corresponds to the GR value. Gravitational wave constraints obtained from data collected up to $2017$, and reported in Ref.~\cite{ligo2019tests} in $2018$, are slightly more stringent than what is possible with very good X-ray constraints obtained from current instruments. The constraining power of both gravitational wave and electromagnetic observations increases in the future, but the former already by $\sim 2026$ will be more stringent than what can be achieved with the latter in $\sim 2034$.}
\label{fig:comparison}
\end{figure*}

We quantitatively explore the above idea to compute the gravitational waves emitted during the early inspiral of a binary system composed of parametrically deformed Schwarzschild BHs. As stated above, we focus on the inspiral regime and thus work in the post-Newtonian framework, in which the two-body problem can be mapped to an effective one body problem, controlled by an effective Hamiltonian that is constructed from the parametrically deformed Schwarzschild spacetime. From this parametrically deformed Hamiltonian, we then compute the binding energy of the binary, and assuming the radiation-reaction force is prescribed as in GR, we then calculate the rate of change of the orbital frequency, and from this the gravitational waves emitted in the frequency domain. We map the result to the parameterized post-Einsteinian framework, and then use the events in the LIGO-Virgo Catalog GWTC-1~\cite{abbott2019gwtc,ligo2019tests}, to place a constraint on the leading-order metric deformation. 

With that at hand, we then redirect our attention to constraints on parametrically deformed metrics with X-ray observations from low-mass X-ray binaries. For this purpose, we simulate and fit observations using the \textsc{relxill\_nk} model~\cite{relxillnk,abdikamalov2019public}. This model employs the formalism of the transfer function for geometrically thin and optically thick accretion disks~\cite{cunningham1975effects} around parametrically deformed black holes. The astrophysical parameters are chosen to represent a typical X-ray binary that may be observed with current X-ray instruments, such as NuSTAR and NICER, and also in the future with ATHENA. This simulated data is then fitted with \textsc{relxill\_nk} to get constraints on the Schwarzschild deformation parameter.

We find that the gravitational wave observations are already placing constraints on parametric deformations of the Schwarzschild spacetime that are slightly more stringent than the constraints that can be placed with very good X-ray observations in the near future with current X-ray detectors. In the left panel of Fig. (\ref{fig:comparison}) we summarize these results, by showing the marginalized posterior and the $90\%$ confidence constraint on the leading-order metric deformation or ``bumpy'' parameter $a_{1}$ for the combined gravitational wave observations of the GWTC-1 catalog and a simulation with current X-ray instruments. 

We also study the constraining power of future observations of these two techniques, by extrapolating our results to the those we will be able to achieve with advanced LIGO (aLIGO) at design sensitivity ($\sim2026$) and with X-ray data from ATHENA ($\sim2034$). In the right panel in Fig. (\ref{fig:comparison}), we present again the marginalized posterior and the 90\% confidence constraint on the $a_{1}$ bumpy parameter for these two future observations, which shows an improvement of roughly an order of magnitude with either of them. As in the GWTC-1 case, observe that the gravitational wave constraint with aLIGO (by $\sim 2026$) will be more stringent than what will be achievable by ATHENA (by $\sim 2034$) at 90\% confidence. Note also that by the time ATHENA flies, gravitational wave constraints should become substantially more stringent than predicted above because of new higher signal-to-noise ratio (SNR) events (expected as the instruments are improved over the next 5 years), as well as due to stacking of several hundreds of sources.

The remainder of this paper presents the details that lead to the conclusions summarized above and it is organized as follows. 
Section~\ref{bumpy} establishes notation and presents the parametrically deformed metric used to characterize deviations to the Schwarzschild solution. 
Sections~\ref{GWs} and~\ref{xrays} present the impact of the deformation parameter on the GW profile and on the X-ray spectrum respectively. In each case we present a brief description of the techniques used, the general framework and the procedure followed to analyze the data. 
Section~\ref{conclusions} provides some final remarks. 
Unless otherwise stated, we use geometric units in which $G=1=c$. 

\section{Parametrically Deformed \\ Black Holes}
\label{bumpy}

In this section, we establish notation by briefly summarizing the parametric deformation of the spacetime that we employ in this work, following Ref.~\cite{Rezzolla:2014mua}, which we hereafter refer to as RZ. The RZ metric is based on a stationary and spherically symmetric spacetime in spherical polar coordinates $(t,r,\theta,\phi)$, where the metric functions are continued fractions in the radial coordinates. These functions are restricted by requiring that the spacetime contain a BH, i.e.,~that the spacetime contain a surface $r = r_{0}$, called the event horizon, where the expansion of radially outgoing null geodesics is zero. Moreover, this metric includes two ``bumpy'' parameters at leading order, $\epsilon$ and $a_{1}$, which control deviations from the Schwarzschild metric. 

The line element of the RZ solution is simple when only the lowest parameters unconstrained by current observational tests of GR are considered~\cite{Rezzolla:2014mua}:
\begin{equation}
ds^{2}=-N^{2}\left(x\right)dt^{2}+\frac{dr^{2}}{N^{2}\left(x\right)}+r^{2}d\Omega^{2},\label{eq:lineelement}
\end{equation}
where $d\Omega^{2}\equiv d\theta^{2}+\sin^{2}\theta \, d\phi^{2}$ is the line element on the two-sphere and $x$ is a compactified radial coordinate defined as
\begin{equation}
x\equiv1-\frac{r_{0}}{r},
\end{equation}
such that $x=0$ corresponds to the location of the event horizon $r = r_{0}$, while $x=1$ corresponds to spatial infinity. The metric functions $N(x)$ is decomposed via
\begin{equation}
N^{2}(x)=xA(x)
\end{equation}
where
\begin{equation}
A(x)=1-(1-x)\left[\epsilon+\epsilon(1-x)-a_{1}(1 - x)^{2}\right]\,.
\end{equation}
The line element shown above is therefore characterized in terms of the bumpy parameters $\epsilon$ and $a_{1}$, which control the magnitude of the non-Schwarzschild deformation, modifying the BH structure near the horizon, and introducing modifications in the spacetime region asymptotically far from the source. 

Let us briefly consider the asymptotic behavior of the metric in these coordinates about the horizon and about spatial infinity. Near the horizon, $r \sim r_{0}$ or $x \ll 1$, we have
\begin{align}
g_{tt} &= -x \left(1 - 2 \epsilon + a_{1}\right) + {\cal{O}}(x^{2})\,,
\nonumber \\
g_{rr} &= \frac{1}{x \left(1 - 2 \epsilon + a_{1}\right)} + {\cal{O}}(x^{2})\,.
\end{align}
This expansion shows that the metric is singular at the location of the event horizon, signaling the presence of a coordinate singularity. Such a singularity renders RZ metric in this coordinate system less than ideal for relativistic numerical simulations. Near spatial infinity, $r_{0}/r \ll 1$ or $x \sim 1$, we have
\begin{align}
g_{tt} &= -1 + \left(1 + \epsilon \right) \frac{r_{0}}{r}  - (\epsilon + a_{1}) \left(\frac{r_{0}}{r}\right)^{3} + {\cal{O}}(r^{-4})\,,
\nonumber \\
g_{rr} &= 1 + \left(1 + \epsilon \right) \frac{r_{0}}{r} + \left(1 + \epsilon\right)^{2} \left(\frac{r_{0}}{r}\right)^{2} 
\nonumber \\
&+ \left( 1 + 2 \epsilon + 3 \epsilon^{2} + \epsilon^{3} + a_{1} \right)
\left(\frac{r_{0}}{r}\right)^{3}
+ {\cal{O}}(r^{-2})\,, 
\end{align}
This shows that the observable, or sometimes called ``gravitational'' mass is simply $2 M = (1 + \epsilon) r_{0}$. 

If one requires that the exterior gravitational field of all massive, stationary and spherically symmetric bodies have the form of Eq.~\eqref{eq:lineelement}, then Solar System observations already place constraints on some of these parameters. The tracking of the Cassini spacecraft requires that the parameterized post-Newtonian parameter $\gamma$ to be satisfy $\gamma - 1 \lesssim 10^{-5}$~\cite{bertotti2003test}. The $\gamma$ parameter is related to the parameters of the above metric via $2 \gamma   M = (1 + \epsilon) r_{0}$, which then implies that $\gamma - 1 = (1 + \epsilon) [r_{0}/(2 M)] - 1 \lesssim 10^{-5}$.  The simplest (though non-unique) way to enforce this constraint is to fix the horizon radius at its Schwarzschild value $r_{0} = 2M$, and then require that $\epsilon \lesssim 10^{-5}$. In particular, X-ray tests of GR typically make this choice, setting $r_{0} = 2 M$ and $\epsilon = 0$, which is also the choice we will make henceforth. With these choices, the spacetime metric is fully determined by the choice of mass $M$ and bumpy parameter $a_{1}$.  

\section{Gravitational Wave Constraints}
\label{GWs}

In this section, we study the impact that a parametrically deformed spacetime has in the generation of gravitational waves by a binary system composed of non-Schwarzschild BHs, i.e.,~composed of objects whose spacetime near either of them approaches the RZ metric. We focus on the early inspiral of the binary system, such that the problem can be studied within the post-Newtonian (PN) framework~\cite{blanchet2014gravitational}, and we work to leading-order in this approximation when considering deviations from GR.

The comparable-mass, two-body problem can be mapped in the PN approximation to an {\it effective one body problem}. In this effective problem, a test particle of mass equal to the reduced mass of the real binary $\mu = m_{1} m_{2}/m$, where $m = m_{1} + m_{2}$ is the total mass and $m_{1,2}$ are the component masses, is in geodesic motion around a BH with mass equal to the total mass of the real binary. In our case, the spacetime exterior to this BH is represented by the parametrically-deformed metric discussed in the previous section. The conservative sector of the orbital motion is then controlled by the effective Hamiltonian~\cite{buonanno1999effective,hinderer2017foundations}, which can be constructed from the contraction of the RZ metric with the four-momenta of the test particle. The effective problem can then be mapped back to the real, comparable-mass, two-body problem to compute the GWs emitted by such a system, when the two BHs are parametrically deformed.

With this strategy in mind, we begin by considering (massive) test-particle motion in an RZ background with total mass $m$ and bumpy parameter $a_{1}$.  The independence of the metric on the time $t$ and angle $\phi$ about the rotation axis implies the existence of two conserved quantities, energy and (z-component of) angular momentum, respectively. For massive particles with rest mass $\mu$, the energy per unit reduced mass (specific energy) $\tilde{E}$ and the specific angular momentum $\tilde{L}$ can be expressed as
\begin{gather}
\tilde{E}=-u_{t}\quad \text{and}\quad \tilde{L}=u_{\phi}, 
\end{gather}
where the four-velocities are given by $u_{t}\equiv {dt}/{d\tau}$ and $u_{\phi}\equiv {d\phi}/{d\tau}$, and $\tau$ is an affine parameter (proper time for massive particles). The equation of motion for $r(\tau)$ can be obtained from the normalization condition $u_{\alpha}u^{\alpha}=-1$ for the four-velocity, namely 
\begin{equation}
\dot{r}=V_{{\rm eff}}=\tilde{E}^2- N^{2} \left(1 + \frac{\tilde{L}^2 }{r^2}\right).
\label{eq:EOM}
\end{equation}
The orbits of the spacetime in Eq.~(\ref{eq:lineelement}) are therefore completely characterized by the values of two orbital parameters, which can be chosen to be $\tilde{E}$ and $\tilde{L}$, as well as the mass parameter $M$ and the bumpy parameter $a_{1}$, and initial conditions. 

A circular orbit is one that satisfies the followings conditions
\begin{equation}
V_{{\rm eff}} = 0=\frac{dV_{{\rm eff}}}{dr}\,.
\label{eq:CondCircular}
\end{equation}
Expanding the effective potential to leading order about small deformations away from Kerr (i.e.,~in $a_{1} \ll 1$),
\begin{equation}
V_{{\rm eff}}=V_{{\rm eff}}^\GR+V_{{\rm eff}}^\RZ  + {\cal{O}}(a_{1}^{2})\,,
\end{equation}
we find that
\begin{equation}
V_{{\rm eff}}^\GR= \tilde{E}^2 +\frac{\left(\tilde{L}^2+r^2\right) (2 m-r)}{r^3}
\end{equation}
and
 \begin{align}
V_{{\rm eff}}^\RZ & =   \frac{8}{r^6} \left[a_{1} m^3  \left(\tilde{L}^2+r^2\right) (r-2 m) \right]\,.
\end{align}
The energy and angular momentum for circular orbits~\cite{wald1984general} can then be found through the conditions $\dot{r}=0$ and $dV_{eff}/dr =0$, which to leading order in small deformations away from Kerr leads to
\begin{align}
\tilde{E} & = \tilde{E}_{GR}+\delta \tilde{E} +  {\cal{O}}(a_{1}^{2})\,,
\label{eq:Egeneral}\\
\tilde{L} & = \tilde{L}_{GR}+\delta \tilde{L} + {\cal{O}}(a_{1}^{2})\,, 
\label{eq:LGeneral}
\end{align}
 where 
\begin{align}
\tilde{E}_{GR} & =  \sqrt{\frac{4m^{2}-4mr+r^{2}}{r(r-3m)}}\, \qquad
\tilde{L}_{GR}  =  \sqrt{\frac{mr^{2}}{r-3m}}.\label{eq:LGR}
\end{align}
and
\begin{align}
\delta \tilde{E} & = \frac{2 m^3 }{r^{5/2}(r-3m)^{3/2}} \left( 2m -r\right) a_{1},
\label{eq:deltaERK}
\\
\delta \tilde{L} & =  - \frac{6 m^{5/2} }{r^2 (-3m+r)^{3/2}}\left(2m-r\right)^2 a_{1}.
\label{eq:deltaLRK}
\end{align}  
A modified Kepler law can be derived from these expressions by expanding $\omega=\tilde{L}/r^2$ in the far field limit
\begin{equation}
\omega^2  =  \frac{m}{r^3} \left[ 1 + \frac{3m}{r}+ \frac{9m^2}{r^2}-\frac{12 m^2}{r^2}  a_{1} + {\cal{O}}\left(a_{1}^{2},\frac{m^{3}}{r^{3}}\right)\right]\,.
\label{eq:ModKepler}
\end{equation}

With the above test-particle considerations in mind, let us now map the effective one body problem back to the real two-body problem. The conservative dynamics of the real two-body problem is described by the total energy $E_{{\rm T}}$, which, for the circular case, can be written in terms of the effective energy $E_{{\rm eff}}= g_{tt}(1+ \tilde{L}^2 /r^2)^{1/2}$~\cite{damour2000determination}  via~\cite{buonanno1999effective}
\begin{equation}
E_{{\rm T}} = m + E_{{\rm b}}=m \left[ 1+ 2 \eta \left( E_{{\rm eff}} -1 \right)\right]^{1/2}\,,
\end{equation}
where we have explicitly separated the rest-mass energy $m$ from the binding energy $E_{{\rm b}}$, and where we have introduced the symmetric mass ratio $\eta = \mu/m$. Expanding to leading order about the GR deformation and to leading order in PN theory, we then find
\begin{equation}
E_{{\rm b}} = E_{{\rm b}}^{\GR}  -\frac{\eta m^{2}}{2r} \left[ 4 a_{1} \left(\frac{m}{r}\right)^{2} + {\cal{O}}\left(a_{1}^{2},\frac{m^{3}}{r^{3}}\right) \right]\,,
\label{eq:H-of-r}
\end{equation}
where $E_{{\rm b}}^{\GR}$ is the binding energy in GR.

The above expression can now be rewritten in terms of the real orbital frequency $F$ of the binary system. This is achieved by noting that the angular frequencies of the real and effective problems are the same, and thus $\omega = 2 \pi F$. With this in mind, we then have that
\begin{equation}
\frac{E_{{\rm b}}\left(F\right)}{\mu} = E_{{\rm b}}^{\GR}\left(F\right)-4a_{1} \left(2\pi mF\right)^{2} + {\cal{O}}\left[a_{1}^{2},(2 \pi m F)^{8/3}\right]\,,
\label{eq:H-of-F}
\end{equation}
where again we work to leading PN order in the GR deformation.

Equations~\eqref{eq:H-of-r} and~\eqref{eq:H-of-F} represent the total energy of the real, comparable-mass, two-body system, with a GR deformation. This modification is proportional to $a_{1}$ and it enters at 2PN order, i.e., it is of ${\cal{O}}(v^{4})$ smaller than the leading PN order term in $E_{{\rm b}}^{\GR}$. In contrast, the 2PN terms in the GR sector (i.e., the terms in $E_{{\rm b}}^{\GR}$ that are of ${\cal{O}}(v^{4})$ smaller than the leading PN order term in this quantity) are not only proportional to the total mass and the symmetric mass ratio, but also to the spin of the bodies. We therefore expect a partial degeneracy between spin terms and the bumpy parameter $a_{1}$. In passing, note here that going to higher PN order in the calculation of the GR deformation would not break this partial degeneracy; terms proportional to spin in the GR deformation will first enter at 1.5PN order higher than the leading PN order at which $a_{1}$ enters, which is 2PN, implying spin-contributions to the $a_{1}$ deformation will enter at 3.5PN order.

The orbital phase for a binary in a circular orbit is given by 
\begin{equation}
\phi\left(F\right)=\int^{F} \left(\frac{dE}{d\omega}\right) \left(\dot{E}\right)^{-1}\omega \; d\omega,
\end{equation}
where $\dot{E}$ is the rate of change of the binding energy of the system due to gravitational wave emission (and emission of any other propagating degree of freedom that may be present in the theory under consideration). From the above expression, it is clear that the gravitational wave phase depends {\it both} on the conservative (time-symmetric) dynamics represented here in the binding energy, as well as on the dissipative (time-asymmetric) dynamics, represented here in the energy loss rate. 

In this work we wish to compare gravitational wave constraints to X-ray constraints, and the latter are only sensitive to the conservative dynamics of geodesic motion around BHs. As noted above, however, GWs are sensitive to both the conservative and the dissipative sectors, so we have to make an assumption on the dissipative sector. In general, one expects that additional (propagating) degrees of freedom, such as dynamical scalar or vector fields in a modified theory, will introduce additional sources of energy and angular momentum loss. In principle, there are three different classes one can identify depending on the PN order at which modifications to the dissipative sector first enter relative to modifications to the conservative sector:  (i) dissipative corrections enter first at a lower PN order than conservative modifications, (ii) they enter first at the same PN order, or (iii) dissipative modifications enter first at a higher PN order than conservative modifications.

We will here focus on deviations that belong to the third and the second class, as they will lead to more conservative constraints. If a theory falls in the first class, for example introducing dissipative corrections at -1PN, 0PN, or 1PN order, then constraints on such GR deviations will be more stringent than what we calculate here. This is because GW observations are better at constraining lower (or negative) PN order effects, as shown theoretically in Ref.~\cite{yunes2009fundamental} and then with data in Ref.~\cite{ligo2019tests}. If a theory falls in the second class, the constraints we find here will not change by more than a factor of order unity, unless the dissipative modification were to exactly (or nearly exactly) cancel the conservative modification; we are not aware of any theory of gravity whatsoever where such perfect cancellation takes place. Finally, if a theory falls in the third class, then the constraints one would be able to place on such a theory will be approximately the same as those obtained without including the higher PN order corrections, as shown in Appendix B of Ref.~\cite{yunes2016theoretical}.

Given these arguments, we here focus only on modifications to the conservative sector, and assume the dissipative sector is {\it not} modified from GR. Therefore, to compute the GR deviation to leading PN order we only need to use the quadrupole formula to leading PN order to model the change of the binding energy via $\dot{E}_{\GR}^{\LO} = -(32/5) \eta^{2}m^{2}r^{4}\omega^{6}$. The evolution of the orbital phase is then
\begin{equation}
\phi(F)=\phi_{\GR}(F)-\frac{25}{4 \eta}\left(2\pi mF\right)^{-1/3}a_{1} + {\cal{O}}\left[a_{1}^{2},(2 \pi m F)^{0}\right]\,, 
\end{equation}
where $\phi_{\GR}(F)$ is the evolution of the orbital phase in GR, which to leading PN order is $\phi_{\GR}^{\LO}(F) =-1/(32\eta)(2\pi m F)^{-5/3}$. 

The correction to the Fourier phase of the GW can now be computed in the stationary phase approximation~\cite{bender2013advanced}, i.e., assuming that its rate of change is much more rapid than the rate of change of the GW amplitude. The Fourier phase can be written as $\Psi_{\GW}(f) =2\phi(t_{0})-2\pi f t_{0}$, where $t_{0}$ is the stationary time defined through the stationary phase condition $F(t_{0})=f/2$, with $f$ the Fourier frequency. Therefore, to leading PN order and to leading order in bumpy deformations, we find 
\begin{equation}
\label{eq:phasedef}
\Psi_{\GW}(f)=\Psi_{\GW}^{\GR}(f) - \frac{75}{8}u^{-1/3}\eta^{-4/5}a_{1} + {\cal{O}}\left[a_{1}^{2},u^{0}\right]\,, 
\end{equation}
where $\Psi_{\GW}^{\GR}(f)$ is the Fourier GW phase in GR, which to leading PN order is $\Psi_{\GW}^{\GR,\LO}(f)=-3/(128 u^{5/3})$, and where $u\equiv\pi \mathcal{M} f$ with the chirp mass $\mathcal{M}\equiv\eta^{3/5}m$. 

Let us map the above GW deformation in the frequency-domain to the parameterized post-Einsteinian (ppE) framework~\cite{yunes2009fundamental}. In the latter, the leading-order modification to the Fourier phase can be written as
\begin{equation}
\Psi_{\GW}=\Psi_{\GW}^{\GR}+\beta u^{b}
\label{eq:ppEYP}
\end{equation}
where $(\beta,b)$ are ppE parameters. Comparing Eq.~(\ref{eq:ppEYP}) with Eq.~(\ref{eq:phasedef}), we find that 
\begin{equation}
\label{eq:ppE-mapping}
b=-\frac{1}{3};\quad \beta=-\frac{75\eta^{-4/5}}{8}a_{1}.
\end{equation}
As already discussed below Eq.~\eqref{eq:H-of-F}, the GR deformation, here characterized by $\beta$, is proportional to $a_{1}$ and also to $\eta$. In particular, note that $\beta$ is independent of the spins of the bodies, since spin corrections will enter at 1.5PN order higher than the leading PN order effect computed here.

The power of the ppE framework is that we can now map constraints from any given GW observation to constraints on particular modifications to GR. To do so, one must first map the ppE parameterization above into the ppE incarnation used by the LIGO collaboration. Following the choices made in the current state of the LIGO software library~\cite{lalsuite}, at $b =-1/3$ (a 2PN correction) the ppE parameter $\beta$ is related to the PN deformation parameter $\delta \phi_{4}$ via
\begin{equation}
\label{eq:ppE-mapping2}
\beta=\frac{3}{128}\phi_{4}\delta\phi_{4}\eta^{-4/5},
\end{equation}
where $\phi_{4}$ is defined to be $\phi_{4}=\left(15293365/508032+27145/504\,\eta+3085/72\,\eta^{2}\right)$~\cite{khan2016frequency}. 

With the above analysis finished, the procedure to place constraints on the bumpy parameter $a_{1}$ from gravitational wave observations is straightforward. First, one must analyze the gravitational wave data collected by the LIGO and Virgo collaboration to find constraints on the parameter $\delta \phi_{4}$; this step is routinely done by the LIGO collaboration itself, with its results made publicly available for example in Ref.~\cite{ligo2019tests}. Then, one must map the $\delta \phi_{4}$ posterior to a posterior on $\beta$ and from that to a posterior on $a_{1}$, or one can equivalently combine Eqs.~\eqref{eq:ppE-mapping} and~\eqref{eq:ppE-mapping2} to find that
\begin{equation}
\label{eq:ppE-mapping-3}
 a_{1} = - \frac{1}{400}\phi_{4}\delta\phi_{4}\,.
\end{equation}
We will constrain the $a_{1}$ deformation parameter using the events reported in the LIGO-Virgo Catalog GWTC-1~\cite{abbott2019gwtc,ligo2019tests} that were found in \emph{both} modeled searches PyCBC~\cite{usman2016pycbc} and GstLAL~\cite{sachdev2019gstlal} with a significance of false-alarm rate (FAR) $<(1000\,\text{yr})^{-1}$, which will lead to conservative constraints. The events in the catalog that satisfy these conditions are GW150914, GW151226, GW170104, GW170608 and GW170814. These events were analyzed with an IMRPhenomPv2~\cite{hannam2014simple,husa2016frequency} model, modified with GR deviations that only represent fractional corrections in the non-spinning portion of each PN phase coefficient, i.e., with a shift in the PN coefficients $\phi_{i}^{\rm no-spin} + \phi_{i}^{\rm spin} \to \phi_{i}^{\rm no-spin} (1 + \delta \phi_{i}) + \phi_{i}^{\rm spin}$ where the superscript represent whether the phase contributions contain spin terms or not. Thus, the modifications reported in Ref.~\cite{ligo2019tests} are only to leading-order and without PN corrections proportional to the spin in the non-GR sector, just as the modification considered here. For a given observation, we infer the marginalized posterior distribution of $a_{1}$ as follows. Given the Markov chain Monte Carlo (MCMC) chains provided by Ref.~\cite{ligo2019tests}, we compute $a_{1}$ at every location of the parameter space that the chains visited, by evaluating Eq.~\eqref{eq:ppE-mapping-3} at the chain's values of $\delta \phi_{4}$ and $\eta$.

\begin{figure}[thb]
\includegraphics[width=\linewidth{}]{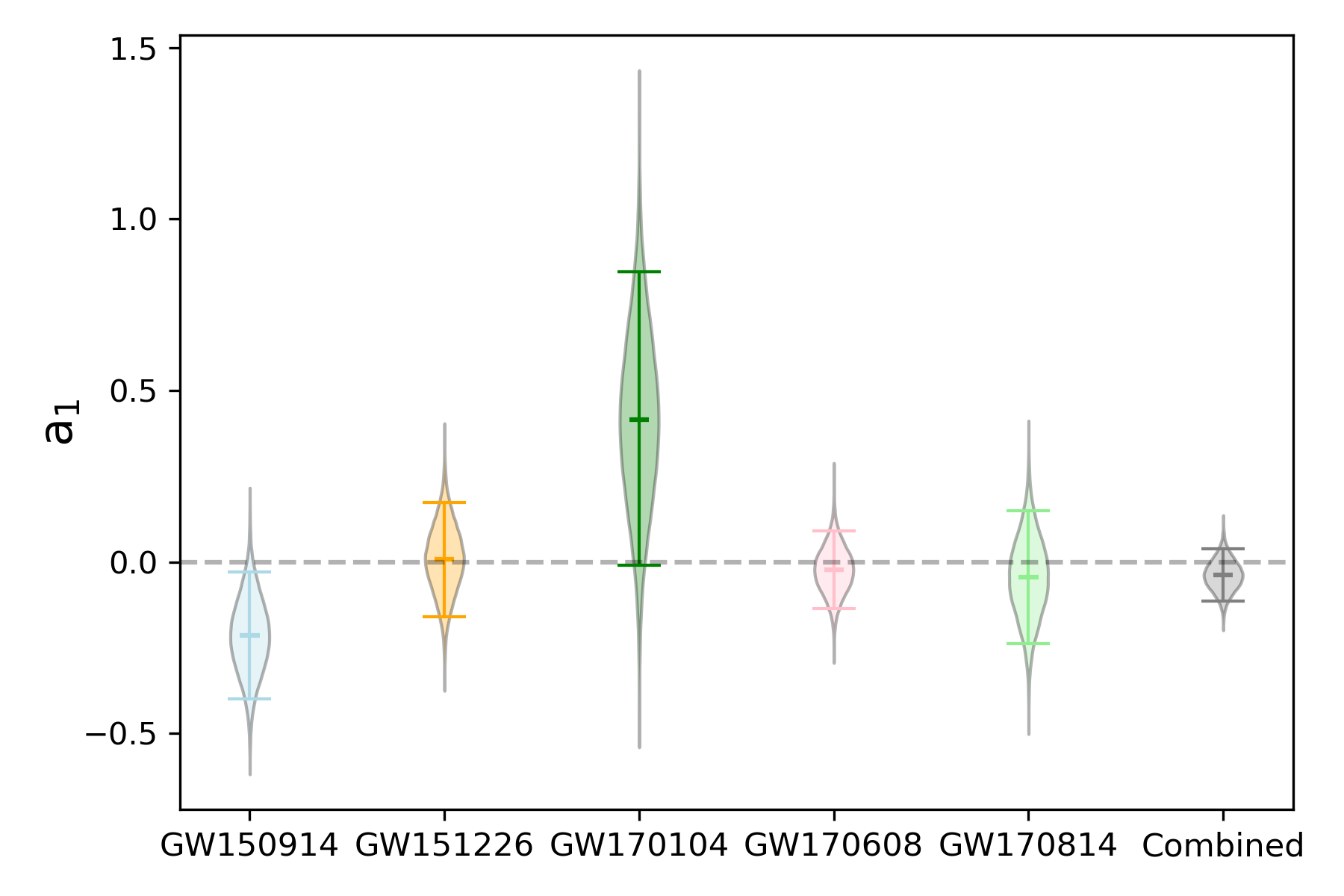}
\caption{Marginalized posteriors on the deformation parameter $a_{1}$ for the most significant binary black-hole events of GWTC-1, and after combining all of these events. The horizontal lines indicate the 90\% credible interval around the indicated mean. The dashed horizontal line at zero corresponds to the GR value.}
\label{fig:GWs}
\end{figure}

Figure~\ref{fig:GWs} shows the posterior on $a_{1}$ obtained from Eq.~\eqref{eq:ppE-mapping-3} using the $\delta \phi_{4}$ and $\eta$ locations in parameter space visited by the MCMC chains, with Table~\ref{tab:combinedvalues} showing the values of the medians and the 90\% credible intervals, for each event. In all cases considered, the posteriors are consistent with the GR value within statistical fluctuations. The likely lightest mass binary BH event, GW170608, gives the strongest constraint because it has a significantly larger SNR in the inspiral regime and it provides many more cycles in the frequency band~\cite{ligo2019tests}. 

\begin{table}[ht]
 \renewcommand{\arraystretch}{1.4}%
\begin{tabular}{m{2.cm}  >{\centering\arraybackslash}m{2.cm} >{\centering\arraybackslash}m{2.cm} }
\hline 
\textbf{Event} & \textbf{$a_{1}$}\\
\hline 
GW150914 & $-0.21_{-0.19}^{+0.18}$\\
GW151226 & $0.01_{-0.16}^{+0.17}$\\
GW170104 & $0.42_{-0.42}^{+0.42}$\\
GW170608 & $-0.02_{-0.11}^{+0.11}$\\
GW170814 & $-0.05_{-0.19}^{+0.19}$\\
\hline 
\end{tabular}
 \caption{\label{tab:combinedvalues} Medians and the 90\% credible intervals on the deformation parameter $a_{1}$ for the most significant binary black-hole events of GWTC-1.}
\end{table}

The bumpy parameters may or may not depend on the parameters of the system. In general, the bumpy parameters will depend on the coupling constants of the modified theory considered, and these are the same for all systems. However, if these constants are dimensionful, the bumpy parameters will also depend on the mass or radius of the objects involved to form a ratio that is dimensionless, as is the case in Einstein-dilaton-Gauss-Bonnet gravity~\cite{metsaev1987order, maeda2009black,Yunes:2011we} and in dynamical Chern-Simons gravity~\cite{alexander2009chern,yunes2009dynamical}. Even if the coupling constants are dimensionless, the bumpy parameters may still depend on dimensionless combinations of the masses, such as the symmetric mass ratio or the dimensionless mass difference. Whether and how the bumpy parameters depend on the system parameters or not will therefore be strongly theory dependent. If the bumpy parameters are independent of the system parameters, we can then enhance our constraints by stacking multiple events. Otherwise, the most stringent constraint can only come from the loudest events. We will now explore both of these cases separately below.

Let us start with the case where $a_1$ is assumed to be independent of system parameters, at least for all BH spacetimes. Given $N$ observations, one can then combine the posteriors on $a_{1}$ through simple multiplication, following e.g., Ref.~\cite{del2011testing}. Given that the GWTC-1 catalog presents a finite number of samples from the posterior distributions relevant to each detection, we have fitted each normalized histogram with a Gaussian distribution. This Gaussian fit is an appropriate approximation to the posterior distribution for each observation, which we can then multiply together in order to get the combined posterior.

Figure~\ref{fig:GWs} presents this combined marginalized posterior, whose mean and $90\%$ confidence region is
\begin{equation}
a_{1}^{\cataT}=-0.038^{+0.075}_{-0.076}\,.
\label{GWConstr}
\end{equation} 
The constraints found above will become more stringent as the statistical uncertainties decrease in the next few years thanks to improvements in detector sensitivity. These improvements lead to some events with very high SNR, and many events with similar SNRs as the events in the GWTC-1 catalog. To conclude this section then, we will estimate the constraining power of future GW observations taking aLIGO at design sensitivity as a benchmark. Let us then assume that by $\sim 2026$ aLIGO at design sensitivity will detect events similar to those in the GWTC-1 catalog studied here. The constraint on deformation parameters with such future events scales as
\begin{equation}
\sigma_{\fut}=\sigma_\cata \left(\frac{N_{\cata}}{N_\fut}\right)^{1/2} \left( \frac{\rho_{\cata}} {\rho_{\fut}} \right),
\label{sigma}
\end{equation} 
where $\sigma$ is the standard deviation of the marginalized posterior (assumed to be Gaussian), $N$ is the number of events detected and $\rho$ is the quadratic mean (or root mean square) of the SNR for all $N$ events. The subscripts ``Obs'' and ``Fut'' denote the derived values found with a current observation and the ones in a future analysis. 

The above relation can be rewritten in a more convenient way by noting that the number of events detected can be expressed as $N = R_{D} \; T \; V$, where $R_{D}$ is the mean intrinsic astrophysical rate of mergers per year per cubic Gpc, $T$ is the number of years of data collected and $V$ is the volume to which the instrument can observe events at a given (threshold) SNR. Using this relation, the standard deviation of the marginalized posterior of future observations scales as 
\begin{equation}
\sigma_{\fut}=\sigma_\cata \left(\frac{T_{\cata}}{T_\fut}\right)^{1/2} \left(\frac{R_{\cata}}{R_\fut}\right)^{3/2} \left( \frac{\rho_{\cata}} {\rho_{\fut}} \right)\,,
\label{sigma2}
\end{equation} 
where $R$ is the range to which the instrument can see at a given (threshold) SNR. Note that the mean astrophysical rate $R_{D}$ has canceled because this is quantity does not depend on the detector used to make observations, but rather on the astrophysics in play during black hole formation and growth.

In order to estimate this quantities we will use the planned sensitivity evolution and observing runs of the aLIGO, AdV and KAGRA detectors over the coming years~\citep{abbott2018prospects}. Advanced LIGO recently finished its second observing run (O2), and started the first half of its third run (O3a) on April 1 2019, which is scheduled to end on September 30 2019 ($T_{\otres}\sim 4\, \text{months}$, as the duty cycle led to double coincidence only for $\sim 80\%$ of the time) with an expected range of $R_{\otres}\sim1200$ Mpc, i.e., an improvement of $1.3$ relative to the range in O2  ($R_{\odos}=910$ Mpc). These improvement implies that the expected number of events during O3a is
\begin{equation}
\left\langle N_{\otres} \right\rangle =N_{\odos}\left(\frac{R_{\otres}}{R_{\odos}}\right)^3  \frac{T_{\otres}}{T_{\odos}}\approx 8,
\label{O3a}
\end{equation} 
where $N_{\odos}=3$ for us (because of the 5 events in the GWTC-1 catalog with a (FAR) $<(1000\,\text{yr})^{-1}$, only 3 were observed in O2), and $T_{\odos} \sim 4$ months of data (O2 lasted for 9 months, but the duty cycle was about $45\%$ for the LIGO network during O2~\cite{abbott2019gwtc}).  The observed events during O3a with very high probability that both components have mass greater than $5 M_{\odot}$ and a FAR $<(1000\,\text{yr})^{-1}$ was $N{\otres}=8$, which shows the estimate above is accurate. 

With this in hand, we can now estimate the strength of projected future combined constraints on bumpy parameters. The three events in O2 led to $\sigma_\odos=0.0582$ and $\rho_{\odos}=14.8$. At design sensitivity the range will increase by roughly a factor of about $2.75$, relative to O2~\cite{abbott2018prospects}. Putting these numbers together, we estimate that aLIGO at design sensitivity should observe approximately $\left\langle N_{\ocinco} \right\rangle \sim374$ events by 2026 (i.e., assuming two years of data collected during O5 at least double coincidence) that are similar to the three events observed in the O2 run. Although most of these $\left\langle N_{\ocinco} \right\rangle$ detections will be found with SNRs close to the detection threshold, assumed here to be $\rho_{\rm {th}} = 12$~\cite{abbott2018prospects},  there will exist a tail of higher SNR events. These are the events that may offer the best constraints on both intrinsic and extrinsic parameters of their sources, as it can be seen in Fig.~\ref{fig:GWs}. Following Ref.~\cite{chen2014loudest} we estimate that for 50\% of the $\left\langle N_{\ocinco} \right\rangle$ events, the loudest one should have a SNR louder than $66$. This is the conservative value we assume for the quadratic mean of the SNR of O5, i.e., $\rho_{\ocinco}=66$. In terms of the constraint quoted above at 90\% confidence, we then obtain
\begin{equation}
a_{1}^{\fut}=0.0^{+0.004}_{-0.004},
\label{GWFConstr}
\end{equation} 
after two years of data collected, where we have chosen to fix the mean at zero here. The right panel in Figure~\ref{fig:comparison} shows the projection of the marginalized posterior on the bumpy parameter $a_{1}$ with aLIGO by 2026. The improvement shown above is conservative for a large number of reasons: (i) by the time O5 takes place, aLIGO will have collected of order $100$ events like those considered here during O3 and O4, (ii) some events during O3, O4 and especially O5 will be at a significantly higher single-detector SNR than those considered here, (iii) O5 will take place with a network of detectors that includes KAGRA and LIGO-India neither of which were included in our conservative estimates.  

Let us now consider the case where $a_1$ depends not just on the coupling constants of the theory, but also on the source properties. In this case, the previous stacking procedure cannot be performed, and instead the most stringent constraint will come from the loudest event. By observing one single event like one of the ones already observed, but with higher SNR, the expression in Eq.~\eqref{sigma} reduces to 
\begin{equation}
\sigma_{\fut}=\sigma_\cata \left( \frac{\rho_{\cata}} {\rho_{\fut}} \right).
\label{sigmaSingle}
\end{equation} 
Out of the $\left\langle N_{\ocinco} \right\rangle$ events estimated above for aLIGO by 2026, following again Ref.~\cite{chen2014loudest}, 0.3\% of the cases (corresponding to one event out of the $374$), should have an SNR louder than $622$, assuming a threshold SNR of $\rho_{\rm {th}} = 12$. By taking the likely lightest mass binary BH event, GW170608, which gave the strongest constraint  with $\sigma_{\rm GW170608}=0.069$ and $\rho_{\rm GW170608}=14.1$, we then find 
\begin{equation}
a_{1}^{\fut}=0.0^{+0.003}_{-0.003},
\label{GWFConstrSingle}
\end{equation} 
This estimate is slightly more stringent that the one obtained by combining multiple observations, i.e., Eq.~\eqref{GWFConstr}. We find that independently of the type of modification that $a_1$ encodes (independent or not of the source parameters), the observations with aLIGO would provide more stringent constraints by 2026 than what we will be able to achieve with future X-ray detectors, such as ATHENA, $\sim 10$ years after O5 has been completed, as we will see in the next section.

\section{X-ray Reflection Spectroscopy Constraints}
\label{xrays}

An important technique to test GR with astrophysical BHs is X-ray reflection spectroscopy. Let us begin by summarizing this technique (for more details, please refer to, for instance, Refs.~\cite{relxillnk,Bambi:2015kza,abdikamalov2019public}). X-ray reflection spectroscopy is based on the so-called \textit{disk-corona} model, where a BH (or any other compact object) is surrounded by an accretion disk (typically assumed to be geometrically thin and optically thick) and a corona. Radiation received from the system is comprised of a thermal component (emitted directly from the disk), a power-law component (thermal radiation emitted from the disk and scattered by the corona), and a reflected component (scattered radiation from the corona reflected from the disk). The thermal component is usually at low energies ($0.1-1$ keV for solar mass BHs, lower still for supermassive BHs) and featureless, compared to the reflected component. For testing GR, the component that has been most studied is the reflected radiation because the fluorescent emission lines are expected to be broadened and skewed when observed far from the source due to a combination of relativistic effects occurring in the strong gravity region. In the pasts few years, an \textsc{xspec} model, called \textsc{relxill\_nk}, has been developed to test GR with reflected radiation~\cite{relxillnk,abdikamalov2019public}. 

We are here interested in studying how well X-ray reflection spectroscopy can be used to test GR. Such a study is not new (see for instance, Refs.~\cite{cao2018testing,xu2018study,tripathi2019constraints,Nampalliwar:2019iti}), but we will repeat it using the same metric parameterization as that used for gravitational waves in the previous section to carry out a fair comparison. The general idea is that we will assume that a certain X-ray instrument has detected a reflected radiation signal and found it consistent with GR. We will then generate a model that includes a parametric deviation in the spacetime and ask how well we can constrain this deformation, given statistical noise. This means that when our deformation parameter in the model is set to zero, then the model matches the simulated data \emph{exactly}. Clearly, this will not happen in reality because our astrophysical models will not be exact representations of the data generated by Nature. Therefore, our procedure here will ignore any systematic errors in modeling, leading to optimistic measures of how well GR can be tested with X-ray observations.  

We focus specifically on the active instruments \textit{NuSTAR} and \textit{NICER}, and the proposed instrument \textit{ATHENA}. \textit{NuSTAR} is a high-energy X-ray telescope in orbit around Earth, operating in the energy band of 3-79~keV, and launched in 2012. \textit{NICER} is a NASA mission, designed as a payload for the International Space Station, and deployed in 2017. We use NICER's science module X-ray Timing Instrument (XTI), operating in the energy band 0.2-12~keV, to simulate the response curve of our synthetic data. \textit{ATHENA} is a future instrument that is currently under development by ESA with a planned launch in 2034. \textit{ATHENA} will provide unprecedented capabilities in terms of angular resolution, effective area, spectral resolution and grasp. We here use \textit{ATHENA}'s planned instrument X-ray Integral Field Unit (X-IFU), which will operate in the band 0.2-12~keV, to simulate our synthetic data. 

We prepare our X-ray simulations in the following way. In order to mimic a \textit{current} observation, we simulate simultaneous observations with \textit{NuSTAR} (with an exposure of 100~ks) and \textit{NICER} (a set of four observations, each of 5~ks exposure). These exposure times and simultaneous observations are typical for the respective instruments (see, for instance, Refs.~\cite{Miller:2018zzw,Ludlam:2018vif,Sanna:2018xrg,Jaisawal:2019adc}). To mimic a \textit{future} observation with \textit{ATHENA}, we simulate a 100~ks observation, which is also expected to be a typical amount of data for that instrument~\cite{Barret:2019dig}. 

\begin{figure}[htb]
\includegraphics[width=\linewidth{}]{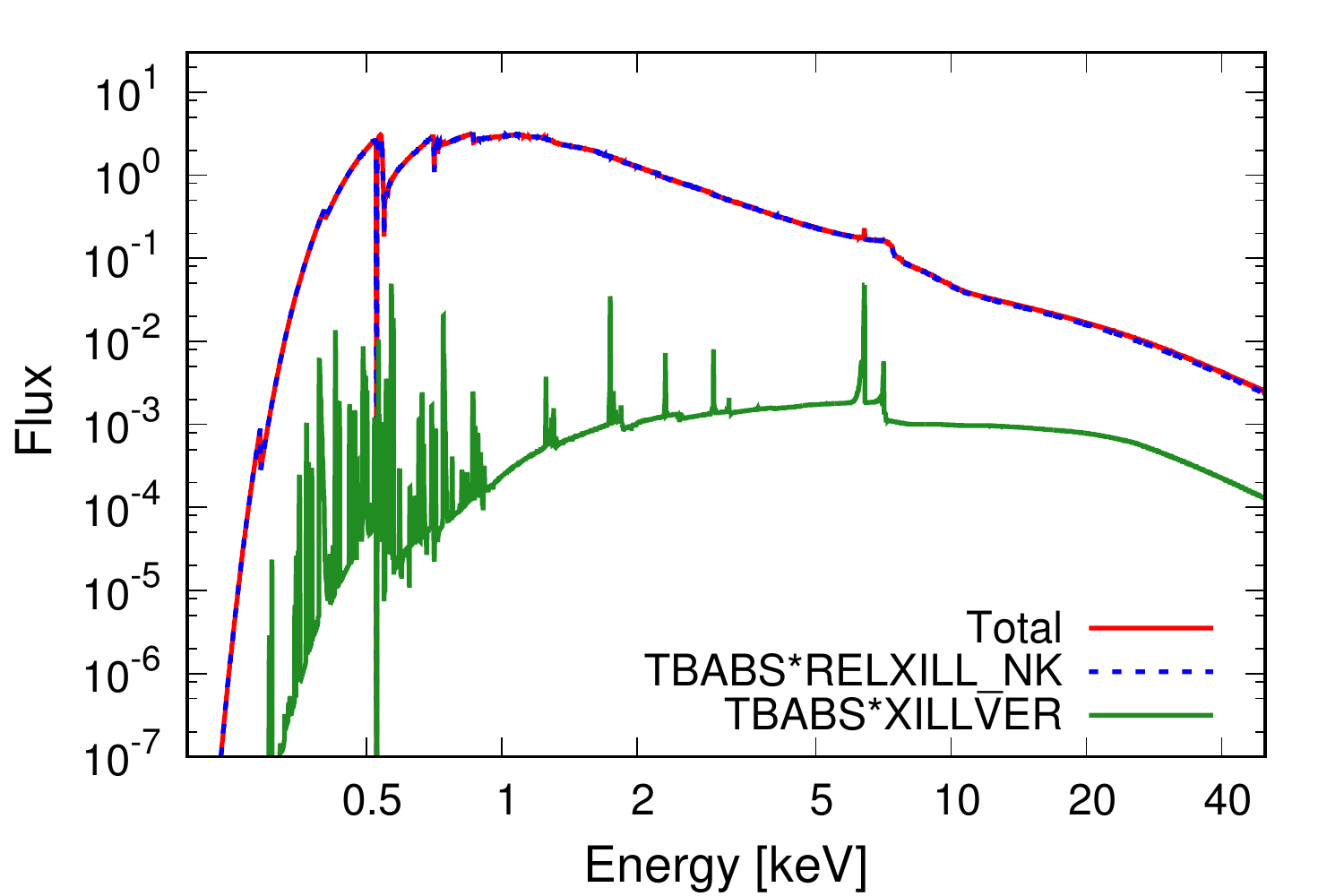}
\caption{(Color Online). Simulated X-ray reflection spectrum. The relativistic reflection spectrum is shown in dotted blue lines, while the non-relativistic spectrum is shown in solid green, and the total spectrum in solid red.}
\label{fig:xrs_spectrum}
\end{figure}

The simulated data (in GR) and the model (outside of GR) are both generated in \textsc{xspec} using 
\begin{center}
\textsc {tbabs*(relxill\_nk+xillver)}.
\end{center}
Here, \textsc{tbabs} is a galactic absorption model, \textsc{relxill\_nk} is our X-ray reflection model and \textsc{xillver} is a model to account for non-relativistic reflection far from the inner regions of the disk, with an example of a simulated spectrum shown in Fig.~\ref{fig:xrs_spectrum}. The metric in Eq.~(\ref{eq:lineelement}) has been implemented in the \textsc{relxill\_nk} framework~\cite{Nampalliwar:2019iti}, including spin and several deformation parameters, although here we choose to work with zero spin and only with one non-zero deformation parameter, $a_{1}$. Note that in Ref.~\cite{Nampalliwar:2019iti} $a_{1}$ is denoted by $\delta_1$. The simulations are generated in \textsc{xspec} using \textsc{fakeit}. The \textsc{fakeit} command creates a spectrum by multiplying the model with the response curve of the instrument and adding a background to it. The simulated data is analyzed using \textsc{xspec}, during which the model parameters are either \textit{frozen} (fixed during the analysis), \textit{tied} (tied to another parameter), or \textit{free} (fitted during the analysis). Frozen parameters are listed in Table~\ref{tab:fixedpars}, while tied (marked by a $\ddagger$) and free parameters are listed in Table~\ref{tab:freepars}. The values chosen for simulating the data are presented in the third column of Table~\ref{tab:freepars}. 

\begin{table}[ht]
  \begin{tabular}{m{2cm} >{\centering\arraybackslash}m{3cm} >{\centering\arraybackslash}m{2.3cm}}
   \hline
   \textbf{Model} & \textbf{Description} & \textbf{Value}\\
   \hline 
    \textsc{tbabs} \\
    $N_H$ [$10^{22}$] & Column density & 0.4 \\ 
    \hline
     \textsc{relxill\_nk} \\
     $a$ & Spin & 0.0 \\
     $r_{\textrm{in}}$ & Inner radius & $r_{\textrm{ISCO}}$\\
     $r_{\textrm{out}}$ & Outer radius & $400M$\\
     $z$ & Redshift & 0\\
     \hline
     \textsc{xillver} \\
     $\log\xi$ & Ionization & 0\\
     $R_f$ & Reflection fraction & -1\\
      $z$ & Redshift & 0\\
     \hline
\end{tabular}
 \caption{\label{tab:fixedpars} List of model parameters which remain fixed throughout the analysis. The radius of the inner most stable circular orbit is denoted by $r_{\textrm{ISCO}}$.}
   
\end{table}

 \begin{center}
 \renewcommand{\arraystretch}{1.3}%
\begin{table*}[ht]

  \begin{tabular}{m{2cm}  >{\centering\arraybackslash}m{3.5cm} >{\centering\arraybackslash}m{2.3cm} >{\centering\arraybackslash}m{2.3cm} >{\centering\arraybackslash}m{2.3cm} >{\centering\arraybackslash}m{2.3cm}}
   \hline
   \textbf{Parameter} & \textbf{Description} & \textbf{Simulated} & \textbf{Current} & \textbf{Future}\\
   $q$ & Coronal emissivity index & 3 & $3.01^{+0.02}_{-0.02}$ & $3^{+\Delta}_{-\Delta}$\\
   $i^{\ddagger}$ [deg] & Inclination & 45 & $44.89^{+0.14}_{-0.14}$ & $45^{+0.01}_{-0.01}$\\
   $\Gamma^{\ddagger}$ & Incident radiation index & 2 & $2^{+\Delta}_{-\Delta}$ &  $2^{+\Delta}_{-\Delta}$\\
   $\log\xi$ & Ionization & 3.1 & $3.1^{+\Delta}_{-\Delta}$ & $3.1^{+\Delta}_{-\Delta}$\\
   $A_{\textrm{Fe}}$ & Iron abundance & 3 & $3.01^{+0.01}_{-0.01}$ & $3^{+\Delta}_{-\Delta}$\\
   $E_{\textrm{cut}}^{\ddagger}$ & Energy cutoff & 300 & $297.43^{+5.71}_{-4.15}$ & $297.59^{+1.60}_{-1.36}$\\
   $R_f$ & Reflection fraction & 1 & $1^{+0.01}_{-0.01}$  & $1^{+\Delta}_{-\Delta}$\\
   $\bf{a_{1}}$ & Deformation parameter & 0 & $-0.07^{+0.080}_{-0.082}$ & $0.00^{+0.007}_{-0.007}$\\
   $N$ [$10^{-3}$]& Norm & 35 & $34.97^{+0.03}_{-0.01}$ & $34.90^{+0.07}_{-0.06}$\\
   \hline
   \textsc{xillver} \\
   $A_{\textrm{Fe}}$ & Iron abundance & 1 & $0.9^{+0.11}_{-0.08}$ & $1.06^{+0.04}_{-0.05}$\\
   $N$ [$10^{-3}$] & Norm & 8 & $8.97^{+0.56}_{-0.75}$ & $7.78^{+0.14}_{-0.12}$\\
   \hline
   \hline
   $\chi^2/dof$ & & & $6169.57/6325$ & $28504.32/28556$\\
   \hline
   \hline
  \end{tabular}
    \caption{\label{tab:freepars} Parameters used in the simulated data of the simulations and their best fit values for different observations. Uncertainties are reported at a 90\% confidence level and rounded off to second decimal place or denoted by $\Delta$, when they were too small. Tied parameters are marked by a $\ddagger$. } 
    
\end{table*}
 \end{center}

\begin{figure}[htb]
\includegraphics[width=\linewidth{}]{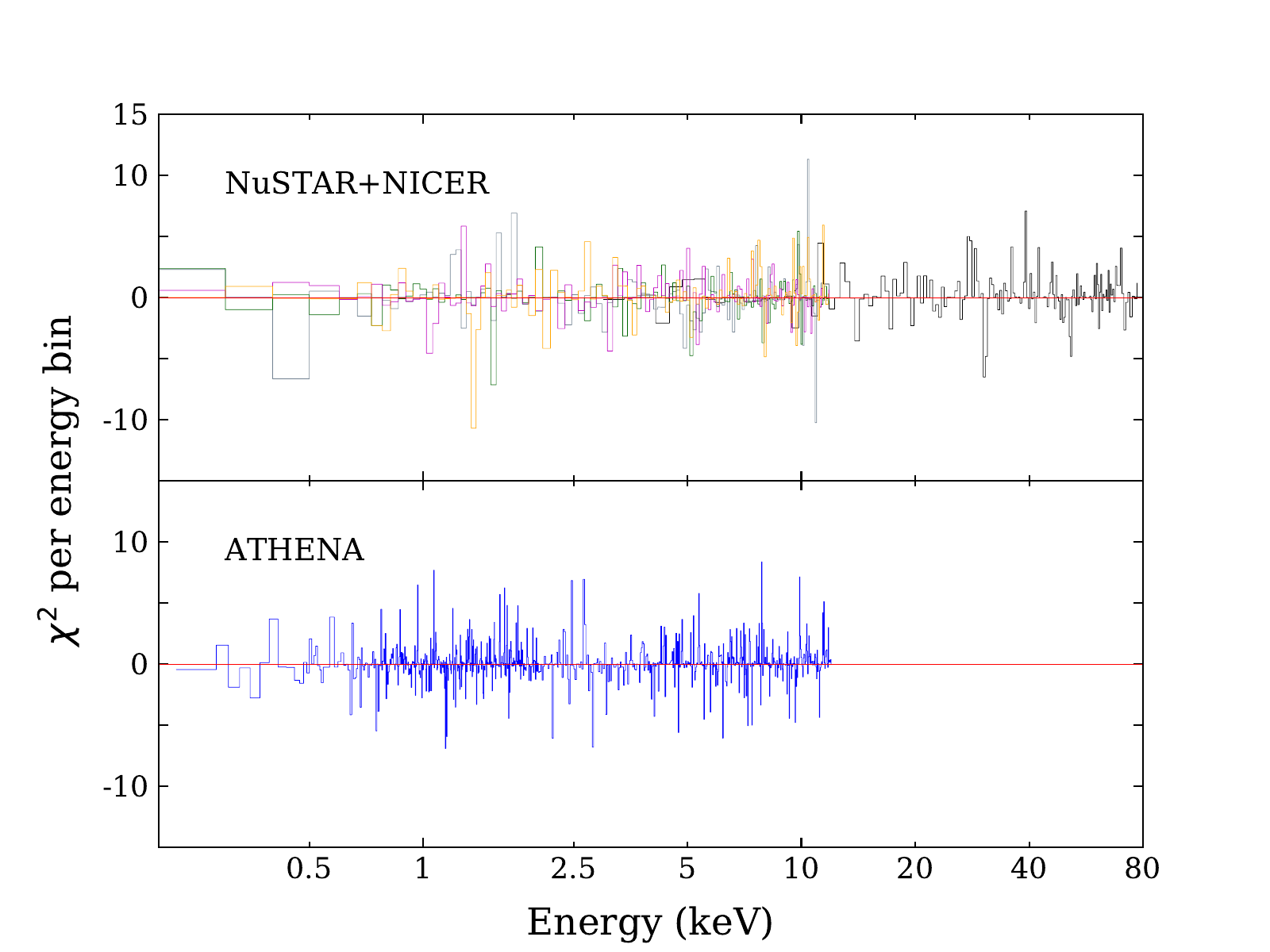}
\caption{(Color Online). $\chi^2$ residuals for best-fit models of each simulation in Table~\ref{tab:freepars}, labeled accordingly. In the top panel, the four NICER simulations (in grey, green, magenta and orange, respectively) and one NuSTAR simulation (in black) are overlaid. The bottom panel shows the case of the ATHENA simulation. The data has been re-binned during plotting for clarity. Observe that there are no unaccounted residuals in the fitted data, confirming that the fits shown in Table~\ref{tab:freepars} are indeed good.} 
\label{fig:xrs_chisq}
\end{figure}

\begin{figure}[htb]
\includegraphics[width=\linewidth{}]{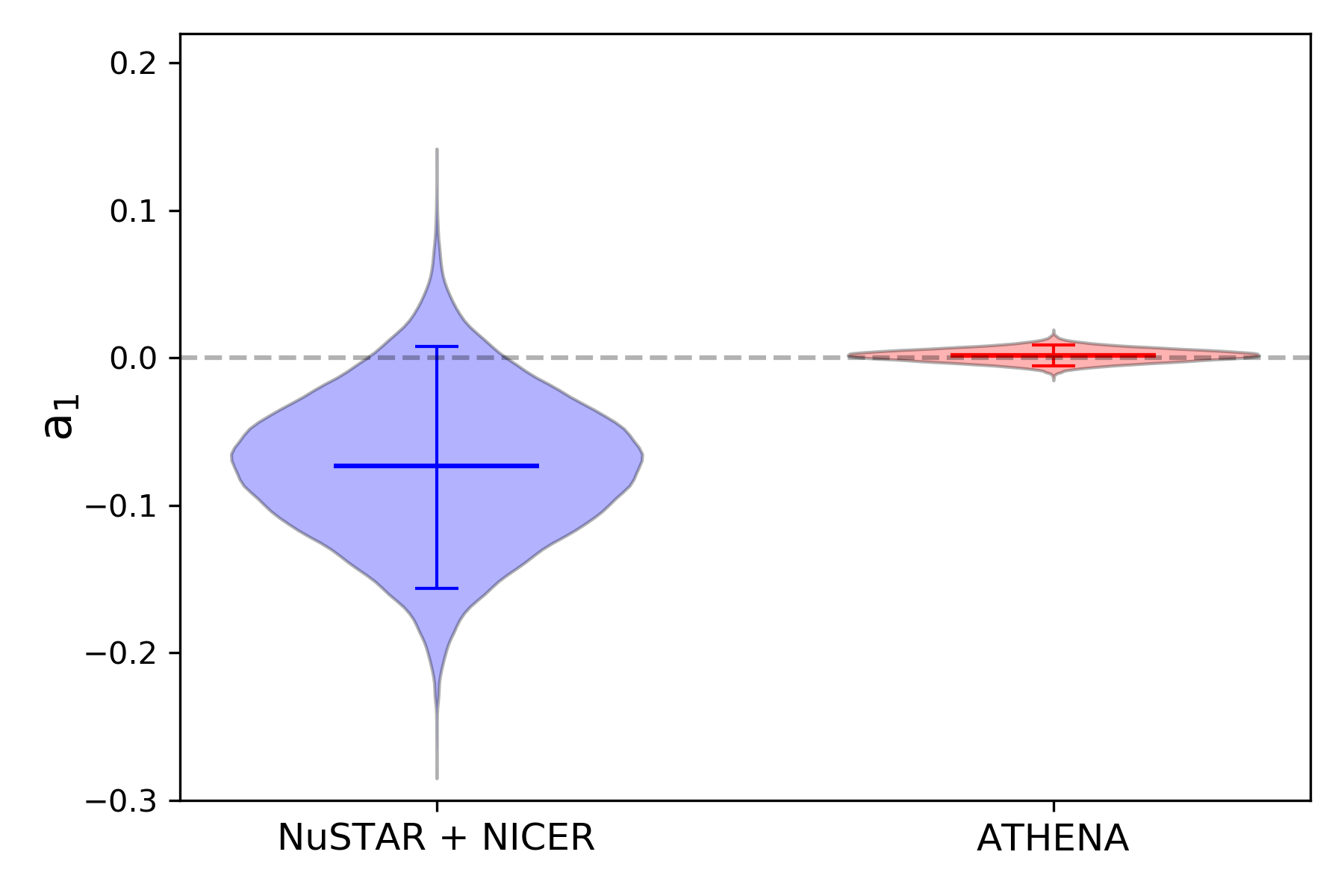}
\caption{Constraints on the deformation parameter $a_{1}$ for the cases presented in Table~\ref{tab:freepars} from simulated observations with combined observations of NuSTAR and NICER, and with ATHENA, respectively, and assuming very good observations of a low mass X-ray binary. The horizontal dashed lines represent 90\% and 99\% confidence levels from the X-ray simulations. Observe that the projected constraints with current instruments are slightly worse than those already placed with GWs. The projected constraints with ATHENA in $\sim 2034$ become one order of magnitude better than those we can place with NICER in the near future, although this improvement is not enough to beat constraints with aLIGO at design sensitivity that can be obtained by $\sim 2026$ (see Fig.~\ref{fig:comparison}).}
\label{fig:xrays}
\end{figure}

The simulated data is analyzed as follows. Starting with default parameter values, the data is iteratively fitted until the reduced $\chi^2$ is close to $1$ (shown in Table~\ref{tab:freepars}), and there are no unexplained residuals (shown in Fig.~\ref{fig:xrs_chisq}). Table~\ref{tab:freepars} also lists parameter uncertainties at 90\% confidence for the different simulated observations. As it can be seen, most of the parameters are recovered well, and their simulated values lie within the range of uncertainty. Our primary goal here, however, is to study projected constraints on $a_{1}$ with current and future instruments. Figure~\ref{fig:xrays} shows the marginalized posterior distribution for the bumpy parameter, constructed assuming a Gaussian distribution from the obtained $\Delta \chi^2$, where
\begin{eqnarray}
\Delta \chi^2 = \chi^2 (a_{1}) - \chi^2_{\textrm{best fit}},
\end{eqnarray}
with $\chi^2 (a_{1})$ calculated at a given value of $a_{1}$ and marginalized over all other free parameters. This figure is produced with the same data as that used in Fig.~\ref{fig:comparison}, although we choose here to present it again to allow for an easier comparison between current and future constraints capabilities with X-ray observations. Figure~\ref{fig:xrays} shows that projected constraints with ATHENA are about one order of magnitude more stringent than projected constraints with NuSTAR and NICER. Note that, as shown in Fig.~\ref{fig:comparison}, ATHENA constraints are slightly weaker than the very conservative projection of what aLIGO at design sensitivity will be able to achieve by around 2026. 

\section{Conclusions}
\label{conclusions}
Testing GR with electromagnetic and gravitational observations have been of interest to a large swath of the physics community over the past few years. These two type of tests, however, have typically been assumed to be disconnected from each other, with one set of observations testing one aspect of gravity theory, and the other set, a different and disconnected aspect. We point out here that this is incorrect, as indeed both sets of observations probe the conservative (time-symmetric) sector of gravitational theories around BHs. Given this, we then do a direct comparison between tests with gravitational wave observations and tests with X-ray observations. 

We find that combined constraints with LIGO/Virgo data during O1 and O2 are slightly better than what could be achieved with current X-ray instruments, even when one ignores systematic errors in the later. Systematic uncertainties in the X-ray measurements, which were not included in our analysis, would only make X-ray constraints on GR even weaker. As aLIGO becomes more sensitive, reaching design sensitivity by $\sim 2025$, the constraints with aLIGO become even more stringent, independently of the nature of the modification. In particular, by $\sim 2026$--$2027$, aLIGO will obtain constraints that would be already more stringent than what future X-ray instruments deployed $\sim 2034$, such as ATHENA, will be able to achieve.

Even though ground-based detectors of gravitational waves place more stringent constraints on GR than electromagnetic observations, the latter technically have access to a larger region of the curvature-potential phase space. Indeed, ground-based gravitational wave detectors are confined to tests in the highest curvatures and potentials possible in Nature. Moreover, the analysis of GW and EM data suffers from different statistical and systematic uncertainties that can make certain effects hard to measure in one and not in the other. For example, EM observations are particularly good at measuring the spin of BHs, while GW observations can only measure a certain projection of the spin angular momentum, and at present, this combination cannot be estimated very accurately. In this sense, EM and GW observations are complementary, in spite of the quantitative difference in the strength of constraints. 

The study of tests of GR with electromagnetic observations carried out here did not include the spin parameter, but we expect that its inclusion will not change the conclusions of our paper. 
In GR, the spin introduces new features in the spacetime, such as frame-dragging and shifts in the location of the event horizon or the innermost stable circular orbit, which then lead to observable consequences in the electromagnetic spectrum. 
When other astrophysical processes that lead to similar effects in the spectrum are properly modeled, then the spin becomes the only parameter that can introduce these new features, allowing one to estimate the spin accurately from data.
A GR deformation to a spinning black hole metric, however, does not typically introduce new observable features in the spectrum that are non-degenerate with other model parameters, such as the spin, the mass or the accretion rate. 
Therefore, the inclusion of spin will not change the conclusions of our paper. 
   
Another future avenue of study is the search for new ways to test GR with EM and GW observations. The usefulness of the approach to test GR employed in this paper, through parametrically deformed metric, is somewhat limited. In order to faithfully represent known modified gravity solutions, many parameters in the deformed metric need to be non-zero, but EM tests in which many parameters are allowed to vary simultaneously become degenerate and uninformative. Furthermore, while some of these parameters will depend on the coupling constants of the particular modified theory, the majority of the parameters will be pure numbers and not necessarily small. The problem is that which parameters depend on these constants depends on the number of constants in the theory, and thus on the particular model considered. Therefore, it is clear that a new method that is more tightly connected to the symmetries (or anomalies) that are being tested or searched for would be highly desirable. 

\begin{acknowledgments}
We thank Javier Garc\'ia, Gabriela Gonz\'alez, Nicholas Loutrel, Leo Stein and Kent Yagi for valuable feedback, and Scott Perkins and Remya Nair for useful comments about data analysis. We also thank Shuo Xin for the development of the \textsc{relxill\_nk} model used here. We also thank the referee, whose feedback enabled us to significantly improve the presentation of our results. A.C.-A. and N.Y. acknowledge financial support through NASA grant 80NSSC18K1352 and NSF grant PHY-1759615. A.C.-A. also acknowledges funding from the Fundaci\'on Universitaria Konrad Lorenz (Project 5INV1). S.N. acknowledges support from the Excellence Initiative at Eberhard-Karls Universit\"{a}t T\"{u}bingen and the Humboldt Foundation.
\end{acknowledgments}


\bibliography{References}

\end{document}